\documentclass[11pt,twocolumn]{aastex701}

\usepackage[utf8]{inputenc}
\usepackage{graphicx}

\received{February 2, 2026}
\revised{March 17, 2026}
\accepted{March 26, 2026}
\submitjournal{the Astrophysical Journal Letters.}

\begin{document}

\title{Formation and disruption of wide binaries in star clusters revealed by $N$-body simulations}

\author[orcid=0009-0003-4608-2611,sname='Bissekenov']{Abylay Bissekenov}
\affiliation{Department of Physics, Xi’an Jiaotong-Liverpool University, 111 Ren’ai Road, Dushu Lake Science and Education Innovation District, Suzhou 215123, Jiangsu Province, People’s Republic of China}
\affiliation{Heriot-Watt University Aktobe Campus, K. Zhubanov Aktobe Regional University, 34 A. Moldagulova Avenue, 030000 Aktobe, Kazakhstan}
\affiliation{Energetic Cosmos Laboratory, Nazarbayev University, 53 Kabanbay Batyr Avenue, 010000 Astana, Kazakhstan}
\email{abylay.bissekenov22@student.xjtlu.edu.cn}  

\author[orcid=0000-0003-3389-2263, sname='Pang']{Xiaoying Pang} 
\affiliation{Department of Physics, Xi’an Jiaotong-Liverpool University, 111 Ren’ai Road, Dushu Lake Science and Education Innovation District, Suzhou 215123, Jiangsu Province, People’s Republic of China}
\affiliation{Shanghai Key Laboratory for Astrophysics, Shanghai Normal University, 100 Guilin Road, Shanghai 200234, People’s Republic of China}
\email{xiaoying.pang@xjtlu.edu.cn}

\author[orcid=0000-0003-2264-7203,sname='Spurzem']{Rainer Spurzem}
\affiliation{National Astronomical Observatories, Chinese Academy of Sciences, 20A Datun Road, Chaoyang District, Beijing 100101, People’s Republic of China}
\affiliation{Astronomisches Rechen-Institut, Zentrum f\"{u}r Astronomie, University of Heidelberg, Mönchhofstrasse 12–14, 69120 Heidelberg, Germany}
\affiliation{Kavli Institute for Astronomy and Astrophysics, Peking University, Yiheyuan Lu 5, Haidian Qu, Beijing 100871, People’s Republic of China}
\email{}

\author[orcid=0000-0002-4601-7065,sname='Shukirgaliyev']{Bekdaulet Shukirgaliyev}
\affiliation{Department of Physics, School of Sciences and Humanities, Nazarbayev University, 53 Kabanbay Batyr Avenue, 010000 Astana, Kazakhstan}
\affiliation{Energetic Cosmos Laboratory, Nazarbayev University, 53 Kabanbay Batyr Avenue, 010000 Astana, Kazakhstan}

\affiliation{Heriot-Watt University, Aktobe Campus, K. Zhubanov Aktobe Regional University, 34 A. Moldagulova Avenue, 030000 Aktobe, Kazakhstan}
\email{}

\author[orcid=0000-0002-0570-7270,sname='Kalambay']{Mukhagali Kalambay}
\affiliation{Heriot-Watt University Aktobe Campus, K. Zhubanov Aktobe Regional University, 34 A. Moldagulova Avenue, 030000 Aktobe, Kazakhstan}
\affiliation{Fesenkov Astrophysical Institute, 23 Observatory street, 050020 Almaty, Kazakhstan}
\affiliation{Faculty of Physics and Technology, Al-Farabi Kazakh National University, Al-Farabi Avenue. 71, 050040 Almaty, Kazakhstan}
\email{}

\author[orcid=0000-0003-4176-152X, sname='Berczik']{Peter Berczik}
\affiliation{Nicolaus Copernicus Astronomical Center, Polish Academy of Sciences, Bartycka 18, 00-716 Warsaw, Poland}
\affiliation{Fesenkov Astrophysical Institute, 23 Observatory str., 050020 Almaty, Kazakhstan}
\affiliation{Main Astronomical Observatory, National Academy of Sciences of Ukraine, 27 Akademika Zabolotnoho Street, 03143 Kyiv, Ukraine}
\email{}

\begin{abstract}

Wide (soft) binaries are expected to be rapidly disrupted in dense stellar environments, yet they are observed in both the Galactic field and open clusters (OCs). In this Letter, we investigate the formation and disruption of wide binaries in star clusters using direct $N$-body simulations. We perform simulations  containing 10,000 objects with varying binary fractions and initial bulk rotation to give an in-depth look into the dynamical evolution of  wide binaries in star clusters. We find that wide binaries dominate early disruption and formation processes during the initial high-density phase of cluster evolution. We propose two semianalytical models to reproduce the evolution of the wide-binary population in simulations.  The exponential model consists of an early, rapid-disruption phase with a time less than 10\,Myr, driven by frequent encounters at high density, and a longer, relaxation-driven phase between 200 and 300\,Myr. The broken power-law model provides break timescales when the decrease of wide binaries slows down during the early and long-term disruption. All timescales from both models agree with each other and decrease with increasing stellar density induced by high primordial binary fraction and cluster rotation. Wide binary disruption is mostly responsible for the early decline in the total binary fraction of the cluster. Such disruption leads to the decrease of radial binary fraction toward the cluster center until 500\,Myr. Our results suggest low-density OCs or stellar groups younger than 10\,Myr as the optimal environments for detecting wide binaries and provide a physical framework for understanding their contribution to the Galactic field population.

\end{abstract}



\section{Introduction} \label{sec:intro}

Most stars are observed in multiple systems \citep{duchene2013, offner2023}. Among them, binary systems are the fundamental components of stellar multiplicity and play a crucial role in regulating the dynamical evolution of stellar systems, such as star clusters.
In dense environments where close encounters are frequent, binaries are commonly classified according to their binding energy: hard binaries, which possess high binding energies and statistically tend to harden through interactions, and soft binaries, which have low binding energies and statistically are progressively softened \citep{heggie1975, hills1975}. Soft/wide binaries have large semi-major axes and small binding energies, which leads to disruption before the relaxation timescale \citep{heggie1975evo}. 

Despite their tendency to disrupt, wide binaries are still observed in the Galactic field and open clusters (OCs). \citet{el-badry2018nov} found more than 53,000 candidates with projected separation of 50--50,000 au within 200\,pc via Gaia DR2. A similar study with Gaia eDR3 \citep{el-badry2023} identified 1.1 million binaries with 99\% confidence, more than 50\% of which have separation $\geq1000$\,au. Past studies suggested that wide binaries observed in the field are dynamically formed and have outlived their natal clusters \citep{kouwenhoven2010, moeckel&bate2010, moeckel&clarke2011, perets&kouwenhoven2012}. 

In star clusters, there are observations of wide binaries in the Orion Nebula cluster (ONC) with a fraction of $\approx5\%$ \citep{jerabkova2019}. Other OCs, such as Alpha Per, the Pleiades, and Praesepe, exhibit a wide binary fraction of $\approx2.1-3\%$  \citep{deacon2020}. The chemical homogeneity of wide binaries supports a common origin and disfavors random pairing \citep{hawkins2020}. Wide binaries may also play an important role in driving stellar collisions and the formation of interacting binaries and merger products \citep{kaib&raymond2014, michaely&peters2020}. 

Previous studies of star clusters have explored a wide range of formation and evolutionary processes. In systems containing binaries, numerous works have investigated how the binary fraction depends on cluster properties and membership \citep{ivanova2005, hurley2007, fregeau2009, bonatto2012, brinkmann2017}. Other studies have focused on cluster rotation and the role of primordial binaries in shaping rotational evolution and angular momentum transport \citep{einsel1999spurzem, spurzem1999, hong2013, mapelli2017, bianchini2018, kamlah2022, livernois2022, tiongco2022, white2025}.
However, most of the aforementioned works included mainly hard binaries and excluded wide binaries. 
Some attempts on soft binary simulations are currently in progress \citep{wu2025,yang2026}.

In this Letter, we are motivated to explore the formation and disruption of wide binaries in the clustered environment by $N$-body simulations. The evolution of wide binaries in star clusters is investigated under various initial conditions, e.g., primordial binary fraction and initial bulk rotation. We aim to quantify the evolution of wide binaries in star clusters via an analytical model to estimate the disruption timescales. 

In Section\,\ref{sec:methods}, we introduce our $N$-body simulations. In Section\,\ref{sec:results}, the results of our simulations of OCs with different primordial binary fractions and rotation rates are presented. We discuss binary disruption/formation in Section\,\ref{sec:bin_hist}, and propose two semianalytical models for wide binary evolution (Section\,\ref{sec:binary-models}). The evolution of binary fraction and its radial dependence are examined in Section\,\ref{sec:bin_frac}. In Section\,\ref{sec:conclusion}, we summarize our results.

\section{$N$-body simulations} \label{sec:methods}

We use \texttt{NBODY6++GPU}, a high-performance GPU-accelerated code for gravitational $N$-body simulations \citep{kamlahnbody, spurzem&kamlah2023}. \texttt{NBODY6++GPU} is a successor of the direct force integration $N$-body codes originally written by Sverre Aarseth \citep{aarseth1985, aarseth1999, aarseth2003, spurzem1999, aarsethtoutmardling2008}. 

Descriptions of the models are shown in Table  \ref{tab:models}. There are 13 sets of $N$-body simulations with different parameters. We use \texttt{McLuster} \citep{kupper2011, kamlahnbody, leveque2022} to generate initial models, and the resulting outputs are used to run all simulations. The density model is based on the King model \citep{kingmodel}, with a concentration parameter ($W_0$) of 6.0. To ensure robustness of the results, we run 5 to 10 models ($n$ in Table \ref{tab:models}) with the same initial conditions but different random seeds to reduce stochasticity. 

 The initial half-mass radius of the models is 2 pc, and the initial eccentricity distribution is thermal \citep{kroupa2008}. Mass ratio $(0.1<q<1.0)$ distribution is uniform for $m>5\,M_\odot$ and random pairing for the remaining \citep{kiminki&kobulnicky2012, sana2012, kobulnicky2014}.  We adopt \citet{kroupa2001} initial mass function with masses ranging between 0.08 and 150 $M_{\odot}$. Our models adopt stellar evolution  \texttt{level C} from \citet{kamlahnbody}, which includes updated stellar evolution settings such as metallicity-dependent winds, delayed-mechanism supernova remnants, etc. The models are terminated just before complete dissolution, when fewer than 100 total stars remain. We set up a tidal field as a point mass galaxy with a mass $\approx1.4\times10^{11}\,M_\odot$, and the clusters are placed at a distance of $\approx8.178$\,\text{kpc} from the Galactic center in circular orbit.

All simulations have the same number of objects ($N_{\rm obj}$) of 10,000, including both singles and binaries. A binary system is considered a single object. Therefore, the number of stars, $N$, differs from the number of objects. The binary fraction ($f_{b}$) is defined as the ratio of the number of binary systems ($N_{b}$) over $N_{\rm obj}$ ($f_{b}=N_{b}/N_{\rm obj}$). We initialize hard and soft primordial binaries with differing setups. 
For hard binaries, the semi-major axis ($a$) is set to follow a uniform distribution in $\rm log(a)$ in the range of $\rm 0-50$ au for $ M<5M_\odot$ and  \citet{sana2012} distribution for $M>5M_\odot$. For soft binaries, we use the right triangular distribution (also known as `flat uniform distribution' in \texttt{McLuster}): $f_x=2(a-a_{\rm min})/(a_{\rm max}-a_{\rm min})$,  where $a_{\rm min}=50$\,au and $a_{\rm max}=50,000$\,au. The choice of this distribution for soft binaries is to maximize the number of wide binaries.
The ratio of hard over soft binaries is $1/4$. We also divide the binaries in our models into three distinct groups by their semi-major axis separation: \textit{close} ($a\leq10$\,\text{au}), \textit{intermediate} ($10<a\leq1000$\,\text{au}), and \textit{wide} binaries ($a>1000$\,\text{au}). 

\begin{deluxetable*}{llllllllllll}
\setlength{\tabcolsep}{3pt}   
\tablewidth{\textwidth}   
\tablecolumns{10}             
\tablecaption{Model parameters and fitted timescales \label{tab:models}}
\tablehead{
\colhead{Category} 
&\colhead{Name} &\colhead{$n$}  & \colhead{N} 
& \colhead{$f_{\rm b,h}$} 
& \colhead{$f_{\rm b,s}$} 
& \colhead{$f_{\rm b,t}$} 
& \colhead{$\omega_0$} 
& \colhead{$t_1$} 
& \colhead{$t_2$} & \colhead{$t_{\rm br1}$} & \colhead{$t_{\rm br2}$} \\
& & & & \colhead{(\%)} &  \colhead{(\%)} & \colhead{(\%)} &  & \colhead{(Myr)} & \colhead{(Myr)} & \colhead{(Myr)} & \colhead{(Myr)}
}
\startdata
Binary  & N12.5k\_05hb\_20sb & 10 & 12.5 k & 5 & 20 & 25 & 0.0 & $7.9 \pm 1.9$ & $314.2\pm71.2$ & $5.9\pm0.8$ & $412.9\pm112.2$ \\
Influence &N15k\_10hb\_40sb & 10 & 15 k & 10 & 40 & 50 & 0.0 & $6.2 \pm 0.9$ & $249.3\pm26.0$ & $4.9\pm0.5$ & $312.9\pm44.7$ \\
(BI)&N17.5k\_15hb\_60sb & 10 & 17.5 k & 15 & 60 & 75 & 0.0 & $5.9 \pm 0.7$ & $238.8 \pm 29.5$ & $4.6\pm0.4$ & $297.3\pm36.7$ \\
&N20k\_20hb\_80sb & 10 & 20 k & 20 & 80 & 100 & 0.0 & $5.9\pm0.8$ & $222.6 \pm 31.8$ & $4.6\pm0.3$ & $275.5\pm36.2$\\
\hline
Rotation &N15k\_10hb\_40sb\_w0\_03 & 5 &  15 k & 10 & 40 & 50 & 0.3 &  $7.5 \pm 1.0$ &	$274.5 \pm 31.5$ & $5.9\pm0.5$ & $341.6\pm34.3$  \\
&N15k\_10hb\_40sb\_w0\_06 & 5 &  15 k & 10 & 40 & 50 & 0.6 & $ 7.4 \pm 1.1$ &	$267.0 \pm 46.2$ & $6.1\pm0.5$ & $343.9\pm48.2$\\
&N15k\_10hb\_40sb\_w0\_09 & 5 & 15 k & 10 & 40 & 50 & 0.9 & $6.6 \pm 0.6$ &	$254.3 \pm 18.4$ & $5.7\pm0.3$ & $350.7\pm45.1$\\
&N15k\_10hb\_40sb\_w0\_12 & 5 &  15 k & 10 & 40 & 50 & 1.2 &$5.5 \pm0.6$ &	$205.8 \pm 41.0$ & $5.1\pm0.6$ & $301.4\pm56.2$\\
&N15k\_10hb\_40sb\_w0\_18 & 5 & 15 k & 10 & 40 & 50 & 1.8 & $4.9 \pm 0.4$ &	$215.1 \pm 32.7$ & $4.6\pm0.2$ & $320.2\pm68.8$ \\
\hline
BI On & N12.5k\_05hb\_20sb\_w0\_12 & 5 &  12.5 k & 5 & 20 & 25 & 1.2 & $5.4 \pm 0.5$ &	$235.9 \pm 65.5$ & $5.0\pm 0.4$ & $353.4\pm105.2$\\
Rotation & N15k\_10hb\_40sb\_w0\_12 & 5 &  15 k & 10 & 40 & 50 & 1.2 &$5.5 \pm0.6$ &	$205.8 \pm 40.8$  & $5.1\pm0.6$ & $301.4\pm56.2$ \\
&N17.5k\_15hb\_60sb\_w0\_12 & 5 & 17.5 k & 15 & 60 & 75 & 1.2 &  $5.1 \pm 0.2$ &	$201.8 \pm 40.1$ & $4.8\pm0.1$ & $310.1\pm42.9$ \\
&N20k\_20hb\_80sb\_w0\_12 & 5 &  20 k & 20 & 80 & 100 & 1.2 &  $4.5 \pm 0.2$ &	$247.5\pm 87.7$ & $4.2\pm0.2$ & $326.0\pm50.5$ \\
\hline
\enddata
\tablecomments{{$n$ denotes the number of models in each set of simulation. $N$ is the initial number of all the stars in each model. $f_{\rm b,h}$ and $f_{\rm b,s}$ are the binary fractions of hard and soft binaries. $f_{\rm b,t}$ is the total binary fraction of hard and soft binaries. $\omega_0$ is the dimensionless angular velocity. 
$t_1$ and $t_2$ are the short and long-term wide binary dissolution timescales of equation (\ref{eq:model}).} $t_{\rm br1}$ and $t_{\rm br2}$ are break timescales when the decrease of wide binaries slows down from equation (\ref{eq:model2})}. 
\end{deluxetable*}

We divide our models into several categories, such as binary influence (BI), rotation, and BI on rotation. In BI models, we investigate the cluster evolution with different $f_b$. For rotating models, we use \textsc{FOPAX} \citep{einsel1999spurzem, kim2002einsel, kim2004lee, kim2008yoon} to generate the 2D orbit-averaged Fokker-Planck initial models that add rotation to $N$-body simulations. Further, we combine them with the \texttt{McLuster} output with proper $N$-body format scaling via the Monte Carlo rejection technique as described in previous similar studies \citep{kamlah2022, bissekenov2025}. 
Rotating King models are parameterized by pair $(W_0,\omega_0)$, where $\omega_0$ is dimensionless angular velocity \citep[Table \ref{tab:models},][]{einsel1999spurzem, kamlah2022, bissekenov2025}. A higher value of $\omega_0$ indicates higher level of rotation.

\section{Results} \label{sec:results}

\subsection{Binary Disruptions/Formations}\label{sec:bin_hist}

\begin{figure}[ht!]
\begin{center}

\includegraphics[width=0.45\textwidth]{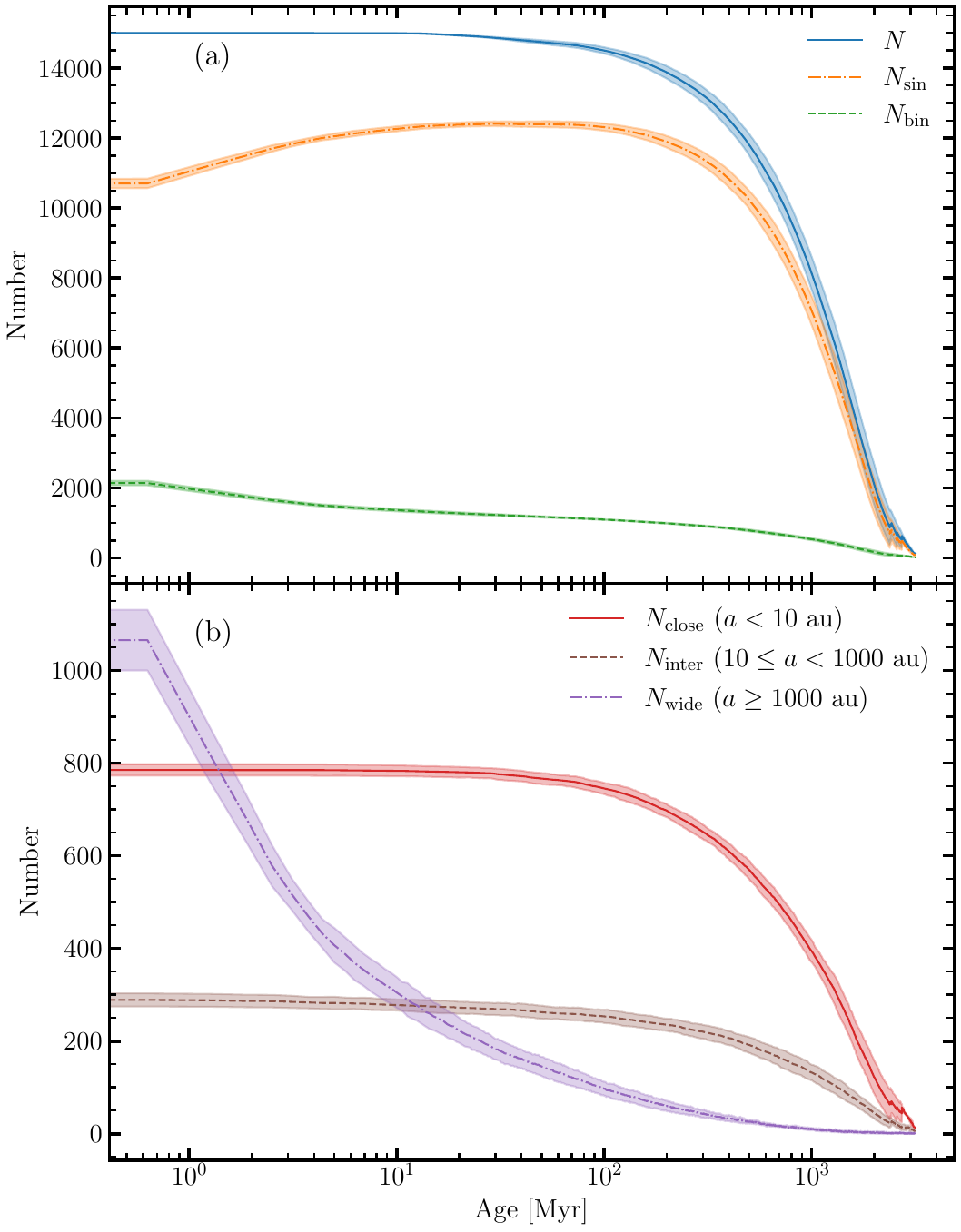} 
\caption{(a): Evolution of total number of stars (blue), single stars (orange), and binaries (green) of the simulation set $N15k\_10hb\_40sb$ from Table \,\ref{tab:models}. (b): Number of close, intermediate (brown), and wide (purple) binaries for the same simulation set. Solid curves and shaded regions represent the mean and standard deviation values among randomized models in the set.}
\label{fig:bin_evol}
\end{center}
\end{figure}

We first look into the evolution of single stars and binaries in the modeled clusters. We use the set of simulations $N15k\_10hb\_40sb$ (Table \ref{tab:models}) as an example to display the results throughout this paper. Figure \ref{fig:bin_evol} shows the evolution of the total number of stars, single stars, and binaries (panel (a)) and the number of close, intermediate, and wide binaries (panel (b)). The simulation is initialized with $f_{b}=50\%$, meaning 5000 binaries, but in panel (a), the initial number of binaries is $\approx2000$. This is due to the instantaneous disruption of very wide binaries caused by the calculations of binding energy in the code. Pairs in disrupted binaries are counted as single stars.

\begin{figure*}[ht!]
\includegraphics[width=\textwidth]{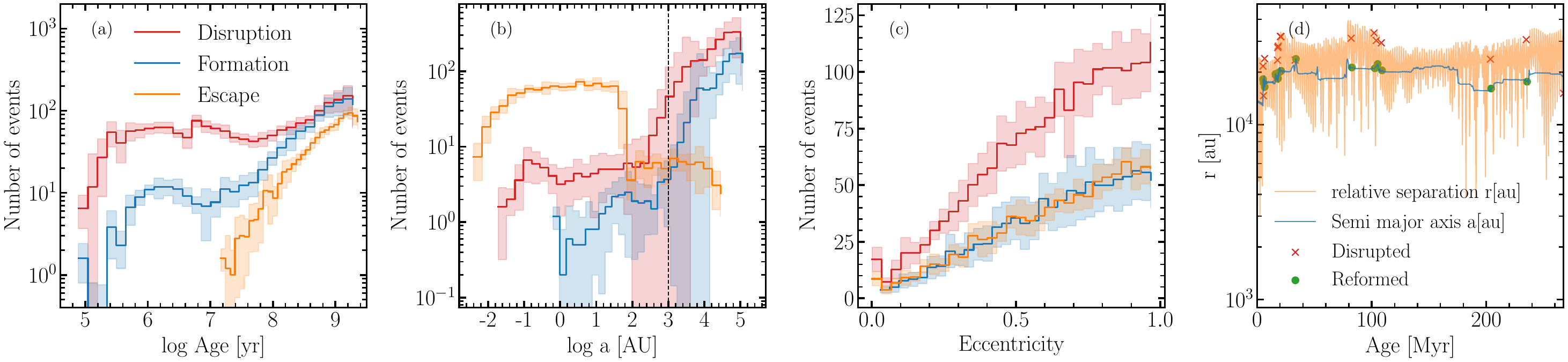} 
\caption{Events of disruption (red), formation (blue), and escape (orange) for all binaries as a function of time (a), distributions of semi-major axis (b), and eccentricity (c) for the set $N15k\_10hb\_40sb$ from Table \ref{tab:models}.  The age axis in (a) is logarithmic; early-time bins span much shorter intervals than those at later ages. Panel (d) shows the evolution of actual separation (orange) and semi-major axis (blue) as a function of time for a specific case of very wide binary with chaotic evolution or highly perturbed by stellar encounters.  Temporary disruption and formation events during the evolution of wide binaries of this type were excluded from (a), (b), and (c).}
\label{fig:bin_hist}
\end{figure*}

To gain a more in-depth understanding of binary disruptions and formations, we record dynamical events in binaries, labeled as \textit{disruption}, \textit{formation}, and \textit{escaper}.  Figure \ref{fig:bin_hist} shows the evolution of the number (panel (a)), semi-major axis distribution (panel (b)), and eccentricity distribution (panel (c)) for the binaries that were disrupted, formed, and escaped during the simulation.
Wide binaries have binding energies very close to zero, making them appear temporarily disrupted (binding energy $>0$) in the internal regularizations of $N$-body codes because of stellar flybys or passages through the cluster center. To ensure physical consistency, we recompute the binding energies for all recorded events and keep only those disruptions with positive binding energy, together with their corresponding formation events. All other temporary (repeated) disruption/formation events are removed from the analysis of the data in panels\,(a), (b), and (c) in Figure\,\ref{fig:bin_hist}, except panel (d), where we show one specific binary experiencing those temporary events. We also do not include instantaneous disruptions at age 0\,Myr.

As shown in panel (a) of Figure \ref{fig:bin_hist}, most disruption and formation events occur at early stages of cluster evolution.
These are the disruptions of primordial soft binaries, which are vulnerable to perturbation due to their wide separation (panel (b) in Figure \ref{fig:bin_hist}). 
As can be seen in panel (b), wide binaries are significantly dominant in the disruption and formation events ($a>1000$\,au), compared to intermediate and close binaries ($a<1000$\,au).
Their evolution is easily affected by encounters, and therefore, they are frequently disrupted.
At the same time, formation follows a trend similar to disruption, but much less efficiently. 
This trend changes at later evolutionary stages, when the number of disruptions becomes comparable to the number of dynamical formations. Some of these late-time formations may be driven by three-body encounters \citep{goodman&hut1993, atallah2024, ginat&peters2024}.
The aforementioned effects collectively explain the overall decline of wide/soft binaries in Figure \ref{fig:bin_evol}. Finally, all event types exhibit a thermal eccentricity distribution in panel (c) of Figure \ref{fig:bin_hist}, as expected from the initially imposed thermal distribution.

As for escapers, they are mostly close binaries, and they start escaping from the cluster after 20 Myr. On the contrary, wide binaries have less chance to escape since the disruption timescale is much shorter than the escaping timescale (panel (b)). These behaviors of binaries are consistent with the studies of \citet{heggie1975, heggie1975evo} and \citet{hills1975}.

We show the actual separation (orange) and semi-major axis (blue) evolution as a function of time for a specific case of a very wide binary that was frequently considered formed and disrupted in panel (d) of Figure\,\ref{fig:bin_hist}. 
Red crosses and green dots show recorded events of disruption and reformation. This kind of behavior can be seen in very wide binaries with $a\geq10\,000$ au. However, their actual separation tends to oscillate. When the separation becomes too large, the binary suffers from a lot of perturbations from neighboring stars, which trigger short-term disruptions and later reformation. A similar evolution of the semi-major axis and relative separation is observed in the work of \citet{stegmann2024}, where such wide binaries are classified as undergoing chaotic evolution or as highly perturbed by stellar encounters. In our simulations, some very wide binaries undergo such chaotic processes rather than experiencing simple monotonic dissolution due to strong perturbations.

\subsection{Wide binary evolution model} \label{sec:binary-models}

By analyzing the behavior of wide binaries presented in Section \,\ref{sec:bin_hist}, we consider two semianalytical models for the evolution of wide binaries in $N$-body simulations.

At the early stage of evolution, stellar density is higher; hence, wide binaries have more frequent encounters that lead to disruptions. On the other hand, frequent encounters can also result in the formation of new wide binaries.  
This somewhat counteracts the disruptions but is less efficient. However, after an initial major decrease, remaining wide binaries continue to decrease, albeit at a slower rate. Therefore, an adequate model explaining wide binary evolution should consist of two parts. 

We first consider a two-part exponential model:

\begin{equation}
    f(t)=a_1e^{-\frac{t}{t_1}}+a_2e^{-\frac{t}{t_2}}
\label{eq:model}
\end{equation}
where $t_1$ is the early disruption timescale, and $t_2$ is the long-term disruption timescale. The coefficients $a_1$ and $a_2$ represent the amplitudes and efficiencies of the early- and long-term disruptions.

Another model we propose is a broken power law:

\begin{equation}
    f(t)=a_1(1+\frac{t}{t_{\rm br1}})^{-\alpha_{1}}+a_2(1+\frac{t}{t_{\rm br2}})^{-\alpha_{2}}
\label{eq:model2}
\end{equation}

where $\alpha_1$ and $\alpha_2$ are the decrease rate of wide binaries, and $t_{\rm br1}$ and $t_{\rm br2}$ are break timescales when the decrease of wide binaries slows down. We find that when we set $\alpha_1$ and $\alpha_2$ as variables, they converge to $\alpha_1=1.5$ and $\alpha_2=2$. Thus, we set them as constants.

\begin{figure}[ht!]
\includegraphics[width=\columnwidth]{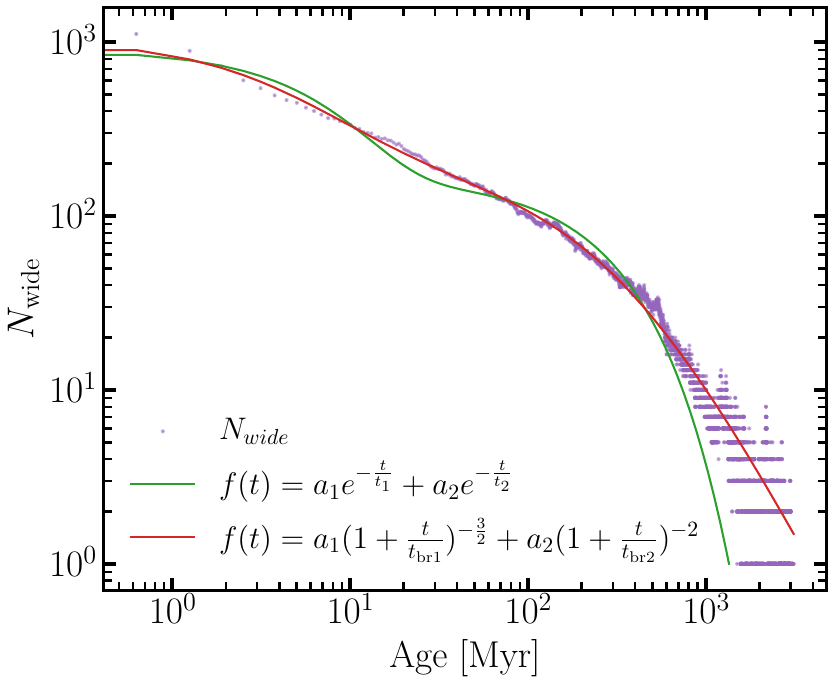} 
\caption{Wide binary evolution models and the evolution of number of wide binaries $N_{\rm wide}$ (purple) in one model from the set $N15k\_10hb\_40sb$ (Table \ref{tab:models}). The green curve is the exponential model (equation (\ref{eq:model})), and the red curve is the broken power-law model (equation (\ref{eq:model2})). }
\label{fig:example}
\end{figure}

We fitted equation\,(\ref{eq:model}) and equation\,(\ref{eq:model2}) for all the models. The fitted equation in comparison to simulations can be seen in Figure \ref{fig:example}, panel (a). 

Performance of the models was quantified using three standard metrics: the coefficient of determination ($R^2$, \citet{montgomery2019applied}), which measures the proportion of variance in the data explained by the model; the root-mean-squared error ($\rm RMSE$, \citet{chai_root_2014}), which reflects the typical magnitude of prediction errors with greater sensitivity to large deviations; and the mean absolute error ($\rm MAE$, \citet{chai_root_2014}), which indicates the average absolute discrepancy between predictions and observations and is robust to outliers.
The exponential and power-law models yield $R^2\approx0.970$ and $0.988$, respectively (values closer to 1 indicate a better fit). Their $\rm RMSE$ values are $\approx7.98$ and $\approx4.96$, and their $\rm MAE$ values are $\approx4.06$ and $\approx1.99$. Both $\rm RMSE$ and $\rm MAE$ are measured in the number of wide binaries.
Given the initial 4000 wide binaries, the values of $\rm RMSE$ and $\rm MAE$ are less than 1\% and therefore acceptable.

Considering the uncertainty in model fitting, the major discrepancy between the exponential model (equation \ref{eq:model}) and simulation occurs at the late stage of cluster evolution due to low number statistics: fewer than 20 wide binaries remain, mostly dynamically formed. Based on previous statistics, the broken power-law model (equation\,\ref{eq:model2}) gives an overall better fit. On the other hand, the exponential model yields meaningful disruption timescales. 
 
The fitted values of $t_1$ and $t_2$ of the exponential model, as well as $t_{\rm br1}$ and $t_{\rm br2}$ of the power-law model, agree with each other, respectively. We present their values for all the simulations in Table\,\ref{tab:models}. 
Wide binary disruptions begin at the crossing time $t_{\rm cross}$, followed by a steady decline until $t_1$ ($t_{\rm br1}$). After that, the disruption rate slows and is driven mainly by encounters from two-body relaxation, decreasing further by $t_{2}$ ($t_{\rm br2}$). A final break occurs around $t_{2}$ ($t_{\rm br2}$), when only $\approx5\%$ of the initially surviving wide binaries remain. The two models complement each other in explaining the wide binary disruption.

Generally, the higher the binary fraction ($f_{\rm b}$), the shorter the disruption timescale.
This is because clusters with higher $f_{\rm b}$ contain more stars and therefore are denser. 
In this kind of dense environment, wide binaries have more severe and frequent encounters that lead to disruptions.

In the rotating models, the faster the cluster rotates (larger $\omega_0$), the quicker the cluster undergoes the gravothermal-gravogyro catastrophe \citep{kamlah2022, bissekenov2025}. The rotating cluster is initialized with high core density and later on forms a bar-like structure. Correspondingly, the disruption timescales $t_1$ and $t_2$ decrease in faster rotating clusters. Therefore, stellar density is the most crucial parameter that influences the disruption of wide binaries.

\subsection{Binary fraction}\label{sec:bin_frac}

\begin{figure*}[ht!]
\begin{center}
\includegraphics[width=\textwidth]{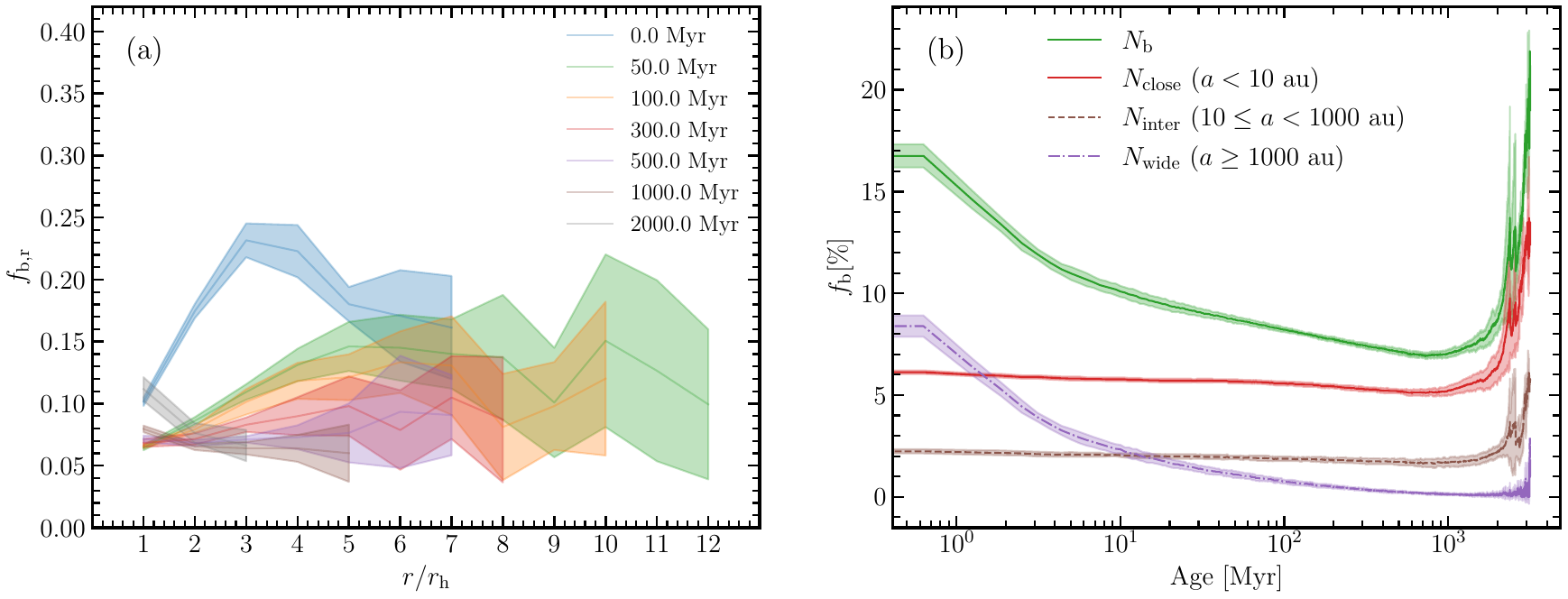} 
\caption{Radial binary fraction of simulations ($f_{\rm b,r}$, a) of the models with $N15k\_10hb\_40sb$ from Table \ref{tab:models}.  Lines represent the mean value of different randomizations, and the shaded region represents the standard deviation. Colors represent ages from 0.0 to 2000 Myr.  (b) shows the binary fraction of all (green), close (red), intermediate (brown), and wide (purple) binaries as a function of time for the same simulation set.}
\label{fig:fb_rad}
\end{center}
\end{figure*}

Figure \ref{fig:fb_rad} panel (a) shows the radial binary fraction for all binary types in the example simulation at different times. The models are initialized as unsegregated, and primordial mass segregation is not present. At 0\,Myr, $f_{b,r}$ drops toward cluster center. This is owing to the disruption of wide binaries at the dense center at the beginning. Wide binaries continue to dissolve up to 4 half-mass radius at the age of 300\,Myr, consistent with the best-fitted long-term disruption timescale $t_2$ (equation\,\ref{eq:model}) of wide binaries (Table\,\ref{tab:models}). Our simulation results show good agreement with observations of OCs in the solar neighborhood \citep{pang2023}, which also found a declining binary fraction toward the center. Around 500\,Myr, the binary fraction along the radial direction is flat, when mostly hard binaries survive. 
In the later stages of evolution, binaries move closer toward the central region \citep{shuqi2021} caused by two-body relaxation. Consequently, the radial binary fraction distribution reaches a peak at the center and declines toward the outer part. 

We present the binary fraction of all, close, intermediate, and wide binaries as a function of time in Figure \ref{fig:fb_rad} panel (b). In our simulations, the close and intermediate binaries are mainly hard binaries. Their fraction changes little over time. The total binary fraction decreases over time, from 17\% to 10\% within 10\,Myr. This decline is mostly due to the disruption of wide binaries, the fraction of which decreases from 8\% at 0\,Myr to 5\% at 2-3\,Myr, and to 2\% at 10\,Myr. Our model resembles ONC in the solar neighborhood, which is observed to contain $\approx5\%$ wide binaries at the age of 2.5\,Myr \citep{jerabkova2019}. Note that our simulations do not include gas expulsion. The cluster initially does not expand significantly, unlike the one undergoing gas removal, as observed in ONC. The expansion lowers the cluster's density and favors the survival of wide binaries. This may explain that even though ONC has 3 times lower number of stars than our modeled clusters, it retains a similar fraction of wide binaries as our models.  In comparison with other much less massive and older OCs like Alpha Per \citep[85\,Myr;][]{navascues2004}, Pleiades \citep[125\,Myr;][]{stauffer1998}, and Praesepe \citep[790\,Myr;][]{brandt2015} from the study of \citet{deacon2020}, the wide binary fraction in our models is $\approx1\%$ lower than theirs. These clusters may experience much more severe gas expulsion and have a lower density. The low-density environment is suitable for the survival of wide binaries, which may become a major contributor to the field population of wide binaries. 

Therefore, the best time to detect wide binaries in an OC is before the early disruption time $t_1$ ($t_{\rm br1}$). Many young filamentary or fractal-shaped stellar groups in the solar neighborhood are young (less than 10\,Myr) and loosely distributed \citep{pang2022}. They can be the best targets for future observations of wide binaries. 

\section{Conclusion} \label{sec:conclusion}

In this Letter, we investigate the dynamical evolution of wide binaries in star clusters using direct $N$-body simulations. Although wide (soft) binaries are expected to be rapidly disrupted in dense stellar environments, observations show that they persist in the Galactic field and are moderately observed in OCs. These motivated our study of formation, disruption, and long-term survival of wide binaries.

We perform 13 sets of \texttt{NBODY6++GPU} simulations with varying primordial binary fractions and initial rotation. Our focus is on the evolution of close, intermediate, and wide binaries in clusters containing 10,000 objects. By tracking binary disruption, formation, and escaping events, we find that wide binaries dominate both disruption and reformation processes, particularly during the early, high-density phase of cluster evolution. Many very wide binaries undergo repeated disruption and reformation due to strong perturbations, exhibiting chaotic behavior rather than simple monotonic dissolution.

Inspired by these results, we adopt a two-component exponential model (equation \ref{eq:model}) and a broken power-law model (equation \ref{eq:model2}), each describing two distinct phases of wide-binary disruption.
The early rapid disruption is driven by frequent encounters at high stellar density, while the longer-term disruption is linked to two-body relaxation. While the power-law model provides a better fit to the simulation data, the exponential model yields physically meaningful disruption timescales. We find that both the exponential disruption timescales and the power-law break times decrease with increasing primordial binary fraction and initial bulk rotation. 
These results indicate that stellar density is the primary factor controlling the survival of wide binaries.

We further examine the evolution of the binary fraction and its radial dependence. The early decline in the total binary fraction is almost entirely driven by the disruption of wide binaries, while hard binaries remain largely unaffected. The resulting wide binary fraction is consistent with observations of the ONC. We find that the radial distribution of the wide binary fraction decreases toward the cluster center, which agrees with observed OCs in the solar neighborhood. Our results indicate that wide binaries are most likely to be detected in young, low-density, and loosely structured stellar systems before the onset of early disruption.

In this work, we provide a coherent physical framework for the evolution of wide binaries in star clusters.  We plan to present a deeper analysis of our simulations in terms of the evolution of compact interacting objects and merger products in the upcoming paper (A. Bissekenov et al. 2026, in preparation).  

\begin{acknowledgments}
We wish to express our gratitude to the anonymous referee
for providing comments and suggestions that helped to improve the quality of this Letter. We thank Jakob Stegmann for the discussions that inspired us to do this work.
This research has been funded by the Science Committee of the Ministry of Science and Higher Education, Republic of Kazakhstan (grant No. AP26102895). X.P. acknowledges the financial support of the National Natural Science Foundation of China through grants 12573036 and 12233013, and the China Manned Space Program with grant No. CMS-CSST-2025-A08. 
P.B. thanks the support from the special programme of the Polish Academy of Sciences and the U.S. National Academy of Sciences under the Long-term programme to support Ukrainian research teams grant No.~PAN.BFB.S.BWZ.329.022.2023. P.B. also gratefully acknowledge the Polish high-performance computing infrastructure PLGrid (HPC Center: ACK Cyfronet AGH) for providing computer facilities and support within computational grant No.~ PLG/2026/019243. 
R.S. acknowledges support from NAOC International Cooperation Office in 2023, 2024, and 2025, from Chinese Academy of Sciences President's International Fellowship Initiative for Visiting Scientists (PIFI, grant No. 2026PVA0089), from the National Natural Science Foundation of China (NSFC) under grant No. 12473017, and from German Science Foundation (DFG) grant Sp 345/24-1.
\end{acknowledgments}

\bibliography{sample701}{}
\bibliographystyle{aasjournalv7}



\end{document}